% Final Version sent to net
\input harvmac
\newcount\figno
\figno=0
\def\fig#1#2#3{
\par\begingroup\parindent=0pt\leftskip=1cm\rightskip=1cm
\parindent=0pt
\baselineskip=11pt
\global\advance\figno by 1
\midinsert
\epsfxsize=#3
\centerline{\epsfbox{#2}}
\vskip 12pt
{\bf Fig. \the\figno:} #1\par
\endinsert\endgroup\par
}
\def\figlabel#1{\xdef#1{\the\figno}}
\def\encadremath#1{\vbox{\hrule\hbox{\vrule\kern8pt\vbox{\kern8pt
\hbox{$\displaystyle #1$}\kern8pt}
\kern8pt\vrule}\hrule}}

\overfullrule=0pt
\def\identity{{\rlap{1} \hskip 1.6pt \hbox{1}}}

\Title{TIFR-TH/00-32}
{\vbox{\centerline{ Fuzzy Cosets and their Gravity Duals}}} 
\smallskip
\centerline{ Sandip P. Trivedi
 \foot{sandip@theory.tifr.res.in} and Sachindeo Vaidya
 \foot{sachin@theory.tifr.res.in}$^,$\foot{Address after Sept. 1st, 2000:
 Department of Physics, University of California, Davis CA 95616, U.S.A.}}
\smallskip
\centerline{  {\it Tata Institute of Fundamental Research,}}
\centerline{\it Homi Bhabha Road, Bombay 400 005, INDIA.}
\smallskip
\bigskip

\medskip

\noindent
$Dp$-branes placed in a certain external RR $(p+4)$-form field expand
into a transverse fuzzy two-sphere, as shown by Myers. We find that by
changing the $(p+4)$-form background other fuzzy cosets can be
obtained.  Three new examples, $S^2 \times S^2$, $CP^2$ and ${SU(3)
\over U(1) \times U(1)}$ are constructed.  The first two are
four-dimensional while the last is six-dimensional.  The dipole and
quadrupole moments which result in these configurations are
discussed. Finally, the gravity backgrounds dual to these vacua are
examined in a leading order approximation. These are multi-centered
solutions containing $(p+4)$- or $(p+6)$-dimensional brane
singularities.

\Date{July 2000}

\newsec{Introduction}
The ideas of noncommutative geometry are finding an increasingly
prominent role in string theory. For example, as has been known for
some time now \ref\wittennon{E. Witten, Nucl.Phys. {\bf B460}, 335-350
(1996), {\tt hep-th/9510135}.}, the coordinates transverse to $N$
$D$-branes do not commute in general and their dynamics is governed by
a non-Abelian gauge theory. More recently, it has been found that
turning on a world volume $U(1)$ gauge field parallel to the
$D$-branes gives rise to a noncommutative version of Yang-Mills
theory.

In \ref\TR{W. Taylor, M. V.  Raamsdonk, ``Multiple Dp-branes in Weak Background Fields",
{\tt hep-th/99100052}.} and  \ref\M{R. C. Myers, "Dielectric Branes", JHEP 9912 022 (1999),
{\tt hep-th/9910053}.},  the T-duality properties of $D$-brane
actions were studied. It was shown that on account of the non-Abelian nature, the
world volume theory for $N$ $Dp$-branes must couple to Ramond-Ramond
(RR) field strengths of degree $p+4$ and above, besides having the
well understood couplings to RR fields of degree $\leq p+2$. The extra
couplings have interesting consequences. Myers showed that for an
appropriate $(p+4)$-form background, the transverse coordinates do not
commute in the ground state and the resulting configuration is
described by a noncommutative generalization of the of the two-sphere,
called the fuzzy sphere. Dispersing the branes in this manner also
results in a dipole moment for the $(p+4)$-form field strength. The
effect is somewhat analogous to the polarization of a neutral atom
when placed in an external electric field: the positive and negative
charges of the atom separate in the external field giving rise to a
dipole moment.

The Myers effect was investigated in the AdS/CFT context by Polchinski
and Strassler \ref\PS{J. Polchinski and M. J. Strassler, "The String
Dual of a Confining Gauge Theory", {\tt hep-th/0003136}.}. They
considered $D3$-branes placed in the corresponding transverse
seven-form field strength background and constructed the supergravity
solution dual to the fuzzy sphere. It was found that the solution
contains a five brane singularity. In fact the resulting spacetime can
be essentially divided into two regions. One, towards the boundary is
approximately the multi-centered $D3$-brane geometry, while the other
deep in the interior, corresponding to the infra-red in the gauge
theory, is the five-brane geometry. The interpolating metric between
these two regions is the gravity background dual to the five brane
with a world volume $U(1)$ field turned on. This establishes, at least
for large 't Hooft coupling, that the infra-red (IR) dynamics of the
$(3+1)$-dimensional fuzzy sphere vacuum is governed by the
$(5+1)$-dimensional five-brane theory with a world volume $U(1)$ field
strength. The two kinds of noncommutativity mentioned at the outset
above, are therefore related under renormalization group flow, in this
system.

The fact that in \M, the transverse coordinates do not commute even in
the ground state, brings the noncommutativity of the geometry seen by
$D$-branes into sharp focus. The purpose of this paper is to study
situations in which this happens in more generality. The particular
background $(p+4)$-form field strength in \M\ is proportional to the
structure constants of $SU(2)$, (and hence preserves an $SO(3)$
subgroup of the $R$-symmetry group) and the resulting configuration is
a noncommutative generalization of the coset $SU(2)/U(1)$. We show in
this paper that more generally, a background which preserves a
subgroup $G$ of the $R$-symmetry group (in the sense that the
$(p+4)$-form field strength is proportional to the structure constants
of the group $G$), gives rise to ground states which are
noncommutative generalizations of particular cosets of $G$. We discuss
which cosets can be realized in this manner and show how coherent
state techniques are useful for analyzing the fuzzy cosets.

In applying these general considerations to string theory we are faced
with a serious limitation: there are too few dimensions to play with!
The dimension of $G$ can at most be the number of transverse
dimensions, which in turn can be no bigger than nine. This allows only
three possibilities, $SU(2)$, $SU(2) \times SU(2)$, and $SU(3)$. The
first, as Myers showed, gives the fuzzy two-sphere. The second, yields
one new surface: fuzzy $S^2 \times S^2$, which is
four-dimensional. The third, gives rise to two cosets of $SU(3)$,
namely $SU(3)/U(2)$ (also known as $CP^2$), and $\displaystyle{SU(3)
\over U(1) \times U(1)}$. These are four- and six-dimensional manifolds
respectively. The resulting configurations acquire a dipole moment
with respect to the $F^{(p+4)}$ field strength.  In addition they
acquire multipole moments with respect to RR field strengths of higher
degree as well. For $S^2 \times S^2$, a quadrupole moment with respect
to $(p+6)$-dimensional field strength; for $CP^2$, a dipole moment
with respect to $(p+6)$-dimensional field strength; and finally, for
$\displaystyle{SU(3) \over U(1) \times U(1)}$, a dipole moment with
respect to the $(p+6)$- and the $(p+8)$-form field strengths. As the
number of $Dp$-branes goes to infinity, the fuzzy surface becomes an
increasingly better approximation to the corresponding manifold, and
one can think of the $Dp$-branes expanding into a higher dimensional
brane which wraps this manifold. In our discussion, the coset manifold
is always embedded in flat space. As a result no net charge is
acquired with respect to the higher dimensional brane.  From the
perspective of the higher dimensional brane, some of the dipole
moments as well as $Dp$-brane charge arises because a topologically
non-trivial gauge field is turned on in the world volume theory.

There is one big difference between the $S^2$ case discussed in \M\
and the $S^2 \times S^2$ and the two cosets of $SU(3)$ discussed in
this paper. In the former case, as is discussed in \PS\ supersymmetry
can be preserved after adding additional mass terms to the theory. In
contrast, in the examples considered here, even allowing for mass
terms, supersymmetry is completely broken. The analysis mentioned in
the previous paragraph is carried out in the tree-level approximation
which is good at weak coupling. What happens at large 't Hooft
coupling is much less certain.

To study this question we turn to the dual supergravity description in
section 5. Our discussion follows \PS\ closely and we adopt the same
strategy of choosing parameters which allow spacetime to be divided
into distinct regions, each governed approximately by one type of
$D$-brane. We find that the gravity backgrounds dual to the fuzzy
cosets mentioned above contain singularities which can be interpreted
as $(p+4)$- or $(p+6)$-dimensional branes.  The presence of
singularities is in accord with expectations based on no-hair theorems
since perturbing the ${\cal N}=4$ theory is dual to adding hair in the
near extremal $p$-brane geometry. The singularities and their dual
descriptions are related to those discussed in \ref\KLT{P. Kraus,
F. Larsen and S.P. Trivedi, JHEP 9903, 003 (1999), {\tt
hep-th/9811120}.}, \ref\KW{I.R. Klebanov and E. Witten, Nucl. Phys. B
556, 89, (1999), {\tt hep-th/9905104}.} \ref\FGPW{D. Z. Freedman,
S. S. Gubser, K. Pilch and N. Warner, {\tt hep-th/9906194}.}
\ref\BS{A. Brandhuber and K. Sfetos, {\tt hep-th/9906021}.}, 
\ref\JPP{C. V. Johnson, A. Peet and J. Polchinski, {\tt hep-th/9911161}.}
\ref\JJ{L. Jarv and C. V. Johnson, {\tt hep-th/0002244}.}. 

Our analysis of the gravity solutions is incomplete in one important
aspect. Take as an example a region of spacetime governed by the
$Dp$-brane metric which crosses over to the $D(p+2)$-brane
geometry. In the solution we construct, we establish that the metrics
in different regions agree in the overlap, but only to leading order
in $(F^{(p+4)})^2/(F^{(p+2)})^2$ - the ratio of two RR field
strengths.  This is not enough, especially in the absence of
supersymmetry. One needs to go to second order at least, before
establishing the existence of the solutions. Unfortunately, the
analysis gets rapidly complicated and we cannot push it this
far. Thus, our discussion of the gravity solutions should be viewed as
only the first step in a more definitive study.

There are many further directions to pursue. It would be revealing to
understand, by an analysis in the $(p+1)$-dimensional gauge theory at
weak 't Hooft coupling, the infra-red dynamics in the fuzzy vacuum, in
particular if it is governed by a $(p+3)$- (or higher) dimensional
theory. Other perturbations of the ${\cal N}=4$ theory, especially
those which preserve supersymmetry and can therefore be controlled
better, are also interesting. Cosets are among the best understood
fuzzy surfaces. However, more general perturbations to the gauge
theory should yield other kinds of noncommutative surfaces as
well. There are close connections between the developments discussed
here and those in \ref\GST{J. McGreevy, L. Susskind and N. Toumbas,
``Invasion of Giant Gravitons from Anti de Sitter Space", {\tt
hep-th/0003075}.}, \ref\LH{P. Ho and M. Li, ``Fuzzy Spheres in AdS/CFT
Correspondence and Holography from Noncommutativity", {\tt
hep-th/0004072}.} which should be pursued in more depth. Finally,
extending this analysis to non-compact groups $G$ might yield examples
of cosmological interest.

One final point. The reader might wonder why we have not considered
other variants of the dielectric effect obtained by turning on a
$(p+6)$- or higher form field strength. In section 4.4 we briefly
discuss one such example which gives rise to a fuzzy generalization of
$S^4$. For the most part though, we postpone a discussion of these
cases for future. This is because, such perturbations of the ${\cal
N}=4$ theory typically result in an unstable theory with runaway
directions in field space along which the energy goes to minus
infinity. For the trilinear terms considered in this paper, runaway
behavior is prevented by the quartic terms present in the ${\cal N}=4$
theory. But for the higher form field strengths which couple to
operators involving more than four scalars, such instabilities are
typically present.  In fact, the fuzzy $S^4$ case mentioned above is
an example of this. It is an extremum of the action but not a minimum:
along a direction in field space the energy goes to minus infinity. We
hope to return to the higher form field strength case in the future.
The runaway behavior could well be absent in the full Born-Infeld
action, or for some specific choices of RR field strengths and other
couplings, which preserve supersymmetry. The instabilities might also
be interesting in their own right and could signify higher dimensional
branes decaying to lower dimensional ones.

Let us end this section by summarizing some additional references.
Two good introductions to some of the ideas in noncommutative geometry
are \ref\Douglas{M. Douglas, "Two Lectures on D-Geometry and
Noncommutative Geometry" {\tt hep-th/99001146}.} and
\ref\Madore{J. Madore, {\it An Introduction to Noncommutative Differential
Geometry and its Applications}, Cambridge University Press, Cambridge,
1995.}. Cosets are discussed from the point of view of coherent states
in \ref\Perelomov{A. Perelomov, {\it Generalized Coherent States and
Their Applications}, Springer-Verlag, Berlin, 1986.} and from the point
of co-adjoint orbits by \ref\kirillov{A. A. Kirillov, {\it Elements of
the Theory of Representations}, Springer-Verlag, Berlin, 1976.}. The
mass deformed ${\cal N}=4$ theory was studied in \ref\DW{R. Donagi and
E. Witten, Nucl. Phys. {\bf B460}, 299 (1996), {\tt
hep-th/9510101}.}. The fuzzy two-sphere was studied in matrix theory in
\ref\taylor{D. Kabat and W. Taylor, Adv.Theor.Math.Phys. {\bf 2},
181-206 (1998), {\tt hep-th/9711078}.}. Field theories on fuzzy $CP^2$
have been studied in \ref\Grosse{H. Grosse and A. Strohmaier,
Lett.Math.Phys. {\bf 48} 163-179 (1999), {\tt hep-th/9902138}.}. There
is a dauntingly large literature on AdS/CFT now, starting with
\ref\maldacena{J. Maldacena, Adv. Theor. Math. Phys. {\bf 2}, 231-252
(1998), {\tt hep-th/9711200}.}, \ref\wittenads{E. Witten,
Adv. Theor. Math. Phys. {\bf 2}, 253-291 , {\tt hep-th/9802150}.} and
\ref\gkp{S. S. Gubser, I. R. Klebanov and A. M. Polyakov,
Phys. Lett. {\bf B428}, 105-114 (1998) {\tt hep-th/9802109}.}, much of
it is well-summarized in \ref\gubserooguri{ O. Aharony, S.S. Gubser,
J. Maldacena, H. Ooguri and Y. Oz, Phys.Rept. {\bf 323}, 183-386
(2000), {\tt hep-th/9905111}.}. The dynamics of $D$-branes with a
world volume $U(1)$ gauge field and its relation to noncommutative
Yang-Mills theory was studied in \ref\cds{ A. Connes, M. R. Douglas
and A.  Schwarz, JHEP 9802 003 (1998), {\tt hep-th/9711162}.} and
\ref\ws{N. Seiberg and E. Witten, JHEP 9909 032 (1999), {\tt
hep-th/9908142}.}. The gravity duals were discussed in
\ref\HI{A. Hashimoto and N. Itzhaki, {\tt hep-th/9907166}},
\ref\MR{J. Maldacena and J. Russo, {\tt hep-th/9908134}.} and
\ref\AJ{M. Alishahiha, Y. Oz and M.M. Sheikh-Jabbari, JHEP 9911 007
(1999), {\tt hep-th/9909215}.}. One recent example of related gravity
solutions is,\ref\PW{K. Pilch, N. Warner, "${\cal N}=1$ Supersymmetric
Renormalization Group Flows from IIB Supergravity '' {\tt
hep-th/0006066}.}. Other gauge theory deformations of interest have
been looked at in \ref\leigh{D. Berenstein, V. Jejjala and
R. G. Leigh, ``Marginal and Relevant Deformations of N=4 Field
Theories and Non-Commutative Moduli Spaces of Vacua'', {\tt
hep-th/0005087}.}.

\newsec{Fuzzy Surfaces}
We start with a brief discussion of fuzzy surfaces and some related
ideas in noncommutative geometry. A readable account of these topics
can be found in \Madore and \Douglas. Here we will settle for a brief
pedestrian account of the subject.

The essential idea behind fuzzy surfaces is that the position
coordinates of the manifold are no longer commuting variables but
instead became operators satisfying an algebra. For example, the
coordinates on classical phase spaces can be thought of as operators
analogous to physical observables in quantum mechanics, and the
algebra they satisfy as being the analog of the algebra of quantum
mechanical observables.

More precisely, a fuzzy surface (see \Douglas) may be defined as a
sequence of algebras $A_N$ which form an increasingly better
approximation to the algebra of continuous functions on some manifold
$X$. Concretely one can think of the algebra $A_N$ in terms of
matrices $M_N$. The eigenvalues of these matrices (and more generally
expectation values of products of matrices) can be compared with
corresponding quantities in the continuous manifold. The $N
\rightarrow \infty$ limit is a classical limit where the expectation
values of the matrices $M_N$ agree arbitrarily well with the
corresponding quantities in the classical manifold.

As an example we consider first the fuzzy $S^2$ surface.  The manifold
$S^2$ can be defined by embedding it in $R^3$ by the relation:
\eqn\defstwo{(X^1)^2 + (X^2)^2 + (X^3)^2=1.}

Let $J^1,J^2, J^3$ be the three angular momentum operators in the spin
$j$ representation of $SU(2)$. These satisfy the relation:
\eqn\cas{J^i J^i = (J^1)^2 + (J^2)^2 + (J^3)^3= j(j+1).}

One can think of the matrices $\hat{X^i} \equiv J^i/\sqrt{j(j+1)}$ as
noncommutative generalizations of the coordinates $X^i$; the relation
\cas can then be identified with \defstwo. It is easy to see that the
limit $j \rightarrow \infty$ is a classical limit: the expectation
values of any product of the matrices $\hat{X^i}$ agrees with the
corresponding quantity in $S^2$ upto corrections of order $1/j$.

The sphere is also a coset $SU(2)/U(1)$. By generalising the
discussion for the sphere, one can construct non-commutative analogues
for some, though not all, coset manifolds.  Let us explain which fuzzy
cosets can be obtained in this manner\foot{ Although many of these
ideas are more general, we will restrict ouselves to compact groups in
this paper.}.  For any compact group $G$ consider a representation $R$
and a weight vector $|\mu>$ in this representation. The isotropy group
$H_{|\mu>}$ of $|\mu>$ is defined to be the subgroup of $G$ which
leaves $|\mu>$ invariant upto a phase. Now for any $G$ consider the
isotropy group $H_{|lws>}$ of the lowest weight
state\foot{Equivalently, one could have chosen the highest weight
state.} in some irreducible representation of $G$.  Then the coset
$G/H_{|lws>}$, for any irreducible representation, can be realised as
a fuzzy surface.  We give a brief argument showing this below.  Before
proceeding let us make three comments. First, note that for $SU(2)$
the isotropy group for the lowest weight vector in any irreducible
representation is $U(1)$.  In general though, depending on the
representation chosen $H_{|lws>}$ can be different.  Second, it is
obvious that $H_{|lws>} $ must always contain the maximal torus of the
group. Third, one can show that the cosets, $G/H_{|lws>}$, are all
even dimensional manifolds with a symplectic form.  In fact they are
complex, homogeneous K\"ahler manifolds \Perelomov.

To obtain a fuzzy generlisation of the coset manifold $G/H_{|lws>}$ we
considers a sequence of irreducible representations, labelled by a
parameter $N$, all of which have lowest weight states with the same
isotropy group $H_{|lws>}$. The dimension of each representation in
the sequence increases as $N$ increases and goes to infinity in the
limit $N \rightarrow \infty$.  Let the generators in the
representation $N$ be denoted as $T^i_N, i=1, \cdots {\rm dim
}\;G$. Then one can show that the limit $N\rightarrow \infty$ is a
classical limit analogous to the $j
\rightarrow \infty $ limit in \cas, and the matrices $T^i_N$ in this
limit describe the coset manifold $G/H_{|lws>}$.

We will briefly sketch an argument which makes this plausible for the
case of fuzzy $S^2$. Since we need to establish that the limit is
classical it is useful to think in terms of coherent states. For other
coset manifolds, the arguments are similar.

As discussed in the appendix, the $SU(2)$ coherent states are of the
form
\eqn\s{ |\xi \rangle = {1 \over (1+|\xi|^2)^j} e^{\xi J_+}
|j, -j \rangle,}
where $|j, -j \rangle$ is the lowest weight vector of the
representation of $SU(2)$ labelled by half-integer $j$, $J_+$ is the
raising operator $J_+=J^1+i J^2$, and $\xi$ a complex number. The
resolution of unity may be written as
\eqn\r{ \int d\mu_j(\xi) |\xi \rangle \langle \xi| = \identity, \quad
{\rm where} \quad d\mu_j(\xi) = {2j + 1 \over \pi} {d^2 \xi \over
(1+|\xi|^2)^2}. } 
This allows us to expand any state in terms of coherent states.
Corresponding to any operator ${\cal O}$ in the Hilbert space, we can
associate a ``classical'' function ${\cal O}(\xi, \bar{\xi}) =
\langle \xi|{\cal O}|\xi \rangle$. For example,
\eqn\e{\langle \xi| \hat{X}^3 |\xi \rangle = - \Bigl ({j \over
\sqrt{j(j+1)}}\Bigr ) {1- |\xi|^2 \over 1+ |\xi|^2}, \quad
\langle \xi| \hat{X}^+ |\xi \rangle = - \Bigl ({2j \over \sqrt{j(j+1)}}\Bigr )
{\bar{\xi} \over 1+ |\xi|^2}, \quad {\rm etc}.  }
It is now immediately clear as to why the $\hat{X}^i$'s go over into the
coordinates in the limit of large $j$: we simply get the stereographic
projection.

The trace of the operator ${\cal O}$ can be calculated in terms of
coherent states as
\eqn\t{ {\rm Tr}\;{\cal O} = \sum_{m=-j}^{j} \langle j,m| {\cal O} |j,m
\rangle = \int d\mu_j(\xi) \langle \xi|{\cal O}|\xi \rangle = \int
d\mu_j(\xi) {\cal O}(\xi, \bar{\xi})\;.}

\newsec{The Dielectric Effect}

By analyzing the T-duality properties of $Dp$-brane actions it was argued 
\TR\ \M\ that RR potentials of degree greater than $p+1$ also couple
to the world volume theory. Let us briefly recall the arguments. Take
the $Dp$-brane world volume to be oriented along the $0,1,\cdots, p+1$
directions. The coupling to the $(p+4)$-form field strength, which
will be the one of main interest in this paper, then takes the form:
\eqn\fform{V_1=-i  \lambda^{-1} {T_p \over 3} \int Tr(X^iX^jX^k)\
F^{(p+4)}_{01\cdots\ pijk}\  dx^0 dx^1 \cdots dx^{p} } 
Here $X^i$ are the scalars transverse to the brane world volume and
are in the adjoint representation of $SU(N)$. The tension $T_p$ of
the $D$-brane and $\lambda$ are
\eqn\consdef{\eqalign{
T_p=&{2 \pi \over g_s (2 \pi \l_s)^{p+1}}, \cr 
\lambda=&2 \pi \l_s^2. }}
In addition the scalar potential for the $X^i$'s has a quartic term
required by ${\cal N}=4$ supersymmetry. Adding it gives a total potential
\eqn\two{V=-{T_p \over 4 \lambda^2} \sum_{a,b} \int d^{p+1} x Tr([X^a,X^b]^2)
 - i  {T_p \over 3\lambda} \int Tr(X^iX^jX^k) 
\ F^{(p+4)}_{01\cdots pijk} \  dx^0 dx^1\cdots dx^{p} }

The first term in \two\ is invariant under the $SO(9-p)$ $R$-symmetry
group of rotations in the $9-p$ transverse directions. If in addition
the $(p+4)$-form RR field strength is a constant and of the form
\eqn\three{F^{(p+4)}_{01 \cdots ijk}= \cases{-{2 \over \lambda} f \epsilon_{ijk}, & 
 for $ i,j,k \in \{ p+1,p+2,p+3 \}$; \cr 0 & otherwise \cr}} 
where $f$ is a real constant, then \two\ preserves a $SO(3) \times
SO(6-p)$ subgroup of the full $R$-symmetry group.

Minimizing \two\ with respect to $X^i$ gives the equations:
\eqn\three{[[X^i,X^j],X^j] + i f \epsilon_{ijk}[X^j,X^k] =0. }
These can be solved by setting
\eqn\four{X^{p+i} = f J^i, \ \   i \in \{p+1,p+2,p+3 \} }
where $J^{i}$ belong to an $N$-dimensional representation of the $SU(2)$
algebra \foot{ The remaining $(6-p)$ scalars must commute with the
three $X^{p+i}$'s in \four\ and with each other.}. Note that in the ground state
the scalars $X^i$ do not commute. In fact, as discussed in the
previous section, $X^i$ represent a noncommutative generalization of
the two-sphere.

The quadratic invariant \foot{The quadratic Casimir invariant $C_r$ in
the $r$th representation of $SU(2)$ is defined by $J^i J^i = C_r
\identity$.} $C_r$ can be used to define the radius $R$ of the fuzzy
two-sphere:
\eqn\rad{X^i X^i \equiv R^2  \identity = f^2 C_r \identity .}
The configuration \four\ has a dipole moment with respect to the
$F^{(p+4)}$ form field strength. From \fform\ we see that the dipole
tensor $P_{ijk}$ is given by
\eqn\defdipole{P_{ijk}=-i \lambda^{-1} T_p \int Tr([X^i,X^j]X^k).}
and is not zero for $i,j,k \in \{p+1,p+2,p+3 \}$ and all distinct.
Thus the externally imposed $(p+4)$-form field strength results in a
dipole moment, analogous to the polarization of a neutral atom placed
in an electric field.

A few comments are worth making about the solution \four. First,
\four\ is different from a multi-centered solution in which
the $N$ branes are uniformly distributed over the two-sphere. In
\four\ the $X^1, X^2$ and $X^3$ coordinates do not commute and hence a 
definite location in all the three directions cannot be simultaneously
assigned to the $Dp$-branes.  A gauge-invariant way to characterize
the difference between the two configurations is the following.  In
the multi-centered solution, the gauge theory is in the Coulomb phase,
while in \four\ it has a mass gap; the heaviest gauge bosons have a
mass $M\sim R/l^2_s$ while the lightest have a mass $R/l^2_s \sqrt{N}$. 

Second, the energy in the minimum \four\ is
\eqn\five{V_N= -{T_p \over 6 \lambda^2}   f^4 N C_r. }
In general, there are several different representations of $SU(2)$
with dimension $N$. The irreducible representation, with the biggest
Casimir invariant, has the largest radius and the lowest energy:
\eqn\six{E_N= -{T_p f^4 \over \lambda^2} {N(N^2-1) \over 24}.}
A reducible representation of the form
\eqn\redrep{X^{p+i}=f J^i_{m \times m} \otimes \identity_{n \times n},\ i=1,2,3}
has  smaller size and higher (but still  negative)  energy:
\eqn\sixa{E_{m}=-{T_p f^4 \over 24 \lambda^2} N (m^2-1).  }
In contrast, the trivial representation, in which all the transverse
scalars have zero expectation values, has zero energy.

Third, in the large $N$ limit the fuzzy surface approximates $S^2$ and
the $Dp$-brane configuration increasingly looks like
$(p+2)$-dimensional branes wrapped on the sphere. One might ask how
the various charges arise from the higher brane's perspective. The
configuration carries no $(p+2)$-brane charge, since the sphere in
question can be embedded in $R^3$ \foot{Locally, there is
$D(p+2)$-brane charge density, but the contributions from the
anti-podal points of the sphere cancel out leaving no net
charge.}. Instead, wrapping on $S^2$ gives rise to a dipole moment. It
turns out that the irreducible representation \six\ corresponds to a
single wrapped $(p+2)$-dimensional brane, while the reducible
representation \redrep\ corresponds to $n$ wrapped $(p+2)$-dimensional
branes. The $Dp$-brane charge arises due to a world volume $U(1)$
magnetic field with components parallel to the two-sphere. This
magnetic field carries magnetic monopole number equal to the number of
$Dp$-branes.  As was mentioned in the introduction, \PS, studied a
supersymmetric version of \three\ for $p=3$, at large 't Hooft
coupling and large $N$.  They found that the low-energy dynamics of
the theory was governed by the $5$-brane theory. It would be
interesting to establish this for small 't Hooft coupling, by a direct
analysis of the gauge theory, in the energy regime,
$R/l_s^2>E>R/(\sqrt{N} l_s^2)$.

Finally, it is worth considering what happens when a mass term of the
form,
\eqn\mass{V_m= {m^2 \over 2} \sum_{i=p+1}^9 (X^i)^2,}
consistent with the $SO(3) \times SO(6-p)$ symmetry, is added to
\two. Putting in the ansatz
\eqn\an{X^{p+i}=a J^i \quad {\rm for} \quad i=1,2,3}
gives the total energy to be
\eqn\tote{V= {T_p  \over 2 \lambda^2} C_r N (a^4 -{4 \over 3} f a^3 +
m^2 \lambda^2 a^2).}
The energy is minimized at
\eqn\mina{a={f+\sqrt{f^2 - 2 m^2 \lambda^2} \over 2}}
and at $a=0$.  For $m^2 \lambda^2$ less than (great than) ${4 \over 9}
f^2$ the energy at the minimum \mina\ is negative (positive) and lower
(larger) than that at the origin.  For large enough mass the
discriminant in \mina\ changes sign and the minimum \mina\ disappears.
Thus the dielectric effect is stable with respect to adding small
enough mass terms in the potential.  The case $m^2 \lambda^2 = {4
\over 9} f^2$ is clearly special. In this case the potential can be
written as a perfect square and the energy at the minimum, \mina, is
zero and equal for all solutions \four.  In fact, adding supersymmetry
preserving mass terms to the ${\cal N}=4$ theory gives rise to this
case \DW.

\newsec{Generalized Dielectric Effect}

\subsec{The General Case}
The essential features of the above discussion are that one started
with an external $(p+4)$-form field strength which preserved an $SO(3)
\times SO(6-p)$ subgroup of the $R$-symmetry.  This gave rise to a
solution which can be interpreted as a noncommutative generalization
of the surface $SU(2)/U(1)$.

We are now ready to generalize this discussion. Start with the
$(p+4)$-form:
\eqn\genf{F^{(p+4)}_{01\cdots pabc}=-{2 \over \lambda} f f_{abc}}
where $f_{abc}$ are the structure constants of some compact group $G$.
The potential \two\ is then minimized when $X^i$ are of the form:
\eqn\gensol{X^{p+i} = f T^{i}, i=1, ...,{\rm dim}\ G,}
where $T^i$ denote the generators of the group $G$ in some
representation of dimension $N$ (the rest of the transverse scalars
are proportional to the identity matrix).
The discussion in section 2 shows that one can always associate a
fuzzy surface with the solution \gensol. 

These fuzzy surfaces are noncommutative generalizations of certain
cosets $G/H$ of $G$. $H$ is determined by the choice of
representation.  For an irreducible representation, $H$ is the
isotropy subgroup of the lowest weight state. Different irreducible
representations can correspond to different isotropy subgroups, $H$,
and thus different cosets $G/H$. As mentioned in section 2, that the
cosets obtained in this manner are all K\"ahler manifolds, and we
denote the K\"ahler form by $K$.

The surface corresponding to \gensol\ has dimension $d= {\rm dim}\ G -
{\rm dim}\ H$.  In general, the configuration \gensol\ carries dipole
moment \defdipole. In addition when $d>2$, dipole moments for higher
degree field strengths $F^{(p+6)}, \cdots F^{(p+d+2)}$ are also
induced; these are defined analogous to \defdipole.

Several features of the discussion for the fuzzy two-sphere carry over
in more generality as well. The quadratic Casimir invariant $C_r$ of
$G$ can be used to assign a "radius" $R$ to the resulting surface:
\eqn\genr{X^i X^i \equiv R^2 \identity = f^2 C_r \identity. }
The vacuum energy can also be expressed in terms of $C_r$ as:
\eqn\genener{E=-{T_p \over 12 \lambda^2} f^4 N C_A C_r,}
where $C_A$ is the Casimir invariant in the adjoint representation of
$G$. Once again we see that the bigger surfaces are also of lower
energy. For large $N$ the fuzzy surface becomes a good approximation
to the manifold $G/H$. The various dipole moments which couple to
$F^{(p+d+2)}, \cdots F^{(p+4)}$ and the $Dp$-brane charge which couples to
$F^{(p+2)}$ can be understood in terms of a $(p+d)$-brane wrapping
$G/H$. For this purpose it is necessary to excite a $U(1)$ gauge field
on the world volume of the $(p+4)$-brane. This field strength is given
by
\eqn\defmag{{\cal F}_{z_i \bar{z_j}} = K_{z_i \bar{z_j}}.}
Finally, the solutions \gensol\ are stable when small mass terms are
added. But there is a critical value for the mass beyond which the
fuzzy surface vacua disappear.

In applying these general considerations to $D$-branes there is one
immediate constraint. There can be at most nine directions transverse
to a $D$-brane. For a compact group $G$ this leaves only three
possibilities: $SU(2)$, $SU(2) \times SU(2)$ and $SU(3)$. The first
choice gives rise to the fuzzy $S^2$ discussed in the previous
section. $SU(2) \times SU(2)$ gives rise to fuzzy $S^2 \times
S^2$. $SU(3)$ yields two cosets $SU(3)/U(2)$ (also known as $CP^2$)
and $\displaystyle{SU(3) \over U(1) \times U(1)}$.

One important remark needs to be made before proceeding further. In
the previous section we mentioned that the trilinear terms giving rise
to the fuzzy $S^2$ can be made supersymmetric after adding appropriate
mass terms. In contrast one finds that, even allowing for mass terms,
the $S^2 \times S^2$ and cosets of $SU(3)$ cases break all
supersymmetries.  While we will not give any details here, one can
verify that even the minimal supersymmetry corresponding to one real
supercharge in $0+1$ dimensions is not allowed in these cases. The
analysis above used the tree-level potential and is valid at weak
coupling. However, in going to the large 't Hooft coupling the absence
of supersymmetry becomes a serious limitation. We will examine this
region of parameter space in the dual gravity description in section
5.

We now turn to discussing the $S^2 \times S^2$ and cosets of $SU(3)$
in more detail.

\subsec{$S^2 \times S^2$}

Here, $G=SU(2) \times SU(2)$, and the $(p+4)$-form field strength \genf\
is :
\eqn\stst{F^{(p+4)}_{01 \cdots pijk}=\cases{-{2 \over \lambda} f_1
\epsilon_{ijk} & if $ i,j,k \in \{p+1,p+2,p+3\}$; \cr 
                                     -{2 \over \lambda} f_2
                                      \epsilon_{ijk} & if $i,j,k \in
                                      \{p+4,p+5,p+6\}$; \cr 
                                      & $0$ otherwise. }}
The solution to \two\ is:
\eqn\ststm{X^{p+i} = f_1 J^{i}_{m\times m} \otimes
                                      \identity_{n\times n}\ ,\ i=1,2,3}
\eqn\ststma{X^{p+3+i} = f_2 \identity_{m\times m} \otimes
J^{i}_{n\times n}\ ,\ i=1,2,3}  
with
\eqn\ststmb{mn=N.}
These represent the fuzzy $S^2 \times S^2$ surface, with radii $m f_1$
and $n f_2$ respectively.  Notice that for given $f_1$ and $f_2$, the
vacua form a one parameter family of solutions labelled by the integer
$m$. The energy of these solutions is:
\eqn\ststenergy{E=-{T_p N \over 24 \lambda^2}  (f_1^4 (m^2-1) + f_2^4 (n^2-1)) }
Depending on the ratio $f_1/f_2$ the lowest energy state has either
$m=1$, or $n=1$ and corresponds to a single fuzzy $S^2$.  When $m f_1
\ll nf_2$ one of the two spheres becomes small and \ststm\ , \ststma\
approach a single fuzzy two-sphere \foot{In fact, in general, another
solution to \two\ is obtained by setting one set of three coordinates,
say $X^{p+i}, i=1,2,3$ to equal \ststm, while the complimentary set of
three coordinates ( and other transverse coordinates) are proportional
to $\identity_{N \times N}$. This configuration corresponds to a
single fuzzy two-sphere.}.

The configuration \ststm, \ststma\ gives rise to a dipole moment for
the $(p+4)$-form. This arises just as in the single $S^2$ case. In
addition there is a quadrupole moment for the $(p+6)$-form field
strength. The quadrupole moment is determined by a coupling of the
form \foot{Strictly speaking a trace term should be removed in the
definition of the quadrupole moment. Imposing the source free
equations for $F^{(p+6)}$ will do this automatically.}:
\eqn\qpole{{T_p \over \lambda^2} \int d^0x\cdots d^px\;
\partial_{n}F^{(p+6)}_{0\cdots pijklm} Tr([X^i,X^j][X^k,X^l] X^m X^n).}  
One can check that for \ststm\ ,\ststma\ this coupling does not
vanish.  From the perspective of a $(p+4)$-brane wrapped on $S^2
\times S^2$ the various charges and moments arise as follows.  A
$U(1)$ field ${\cal F}$ is turned on in the world volume of the
$(p+4)$-brane. As mentioned in the previous section ${\cal F}=K$, the
K\"ahler form for $S^2 \times S^2$. The $Dp$-brane charge is
determined by $\int_{S^2 \times S^2} {\cal F} \wedge {\cal F}$, and
the $D(p+2)$-brane dipole moment by $\int_{S^2} {\cal F}$ on the two
$S^2$'s. One finds that the solution \ststm\ ,\ststma\ corresponds to
$n$ $D(p+2)$-branes wrapping the first $S^2$ and $m$ $D(p+2)$-branes
wrapping the second $S^2$.  To see how the quadrupole moment arises
one can think of wrapping the two $S^2$'s in turn. Wrapping the
$(p+4)$-brane on the first $S^2$ gives rise to a dipole moment for the
$F^{(p+6)}$ form. Wrapping further on the second $S^2$ gets rid of
this dipole moment but generates a quadrupole moment instead.

\subsec{Cosets of $SU(3)$}
\leftline{\it {4.3.1 \ \  General Features}}
We turn next to the case obtained by taking $G=SU(3)$. Since we need
at least eight transverse coordinates, this can be realized only in
the $D1$- or $D0$-brane theory. The solution \gensol\ corresponds to
taking eight of the transverse coordinates to be in an $N$ dimensional
representation of $SU(3)$. Irreducible representations of $SU(3)$ are
parametrized by two integers $(n, m)$ (corresponding to the number of
fundamental and anti-fundamental indices).  One can show that the
lowest weight vector in the completely symmetric representation,
$(m,0)$ or $(0,m)$ has an isotropy group $H=U(2)$ \foot{It is clear
that any vector in the fundamental representation has a $U(2)$
isotropy group. It then follows that the lowest weight vector in the
symmetric representation of $m$ anti-fundamentals must also have the
same isotropy group.}.  Thus when $X^{p+i}$ as defined in \gensol\ are
in the representation $(m, 0)$ or $(0, m)$, one gets the
noncommutative version of $CP^2$. For all other $(n, m)$ one can show
that the isotropy group is the maximal torus $T^2$ of $SU(3)$,
resulting in fuzzy $\displaystyle{SU(3) \over U(1) \times
U(1)}$. Reducible representations correspond to taking a (disjoint)
union of surfaces.

Before proceeding let us clarify one point. Strictly speaking, as was
discussed in section 2, a fuzzy surface corresponds to a sequence of
representations. Any irreducible representation $(n, m)$ can be
regarded as an element of a sequence where $n/m$ is kept fixed and $m
\rightarrow \infty$.  In our discussion above, we have implicitly
assumed such a sequence.

We now discuss the two cosets in some more detail.  The representation
$(n, m)$ has dimension
\eqn\dimsut{D(n,m)= {(n+1) (m+1) (n+m+2) \over 2}.}
Choosing the symmetric representation $(0, m)$, and setting the total
dimension equal to $N$ gives $m^2 \sim 2N $ for large $N$. The Casimir
of this representation can be calculated to be
\eqn\cassym{C_{(0, m)} \simeq m^2/3 \simeq 2N/3.} 
This gives, on setting $C_A=3$ in \genener\  an energy for the $CP^2$ surface,
\eqn\syme{E_{(0, m)} \simeq -{T_p \over \lambda^2}{f^2 \over 12} N m^2 \simeq
-{T_p \over \lambda^2}{f^2 \over 6}N^2,}
and a radius
\eqn\rad{R^2 \simeq {2\over3} f^2 N.} 
for large $N$.

For the representation $(n, m)$, in the limit of large $N$ with $n/m$
fixed, we get that $n, m \simeq N^{1/3}$.  The corresponding Casimir
$C_{(n, m)} \sim N^{2/3}$ leading to an energy,
\eqn\genener{E_{(n,m)} \sim -{T_p f^4 \over \lambda^2} N^{5/3},}
and a radius
\eqn\genradius{R^2 \sim f^2 N^{2/3}.} 

From \syme\ and \genener\ we see that the $CP^2$ surface has the
lowest energy and the the largest size. The energy is lower by a power
of $N^{1/3}$, while the radius is larger by a power of $N^{1/6}$ for
large $N$.  Reducible representations all have an energy which is
higher than the symmetric representation $(m,0)$. For example, the
reducible representation containing $k$ copies of the symmetric
representation has energy
\eqn\eneredsym{E \simeq -{T_p \over \lambda^2}{f^4 \over 6} {N^2 \over
k}.} 

In Appendix B and C, we discuss how the K\"ahler form for the two
manifolds can be calculated from the appropriate representations using
coherent state techniques. The metric of $CP^2$ is the well known
Fubini-Study metric and is a generalization of the round metric for
$S^2$. It has only one free parameter, the overall scale which is
fixed by the radius R. The metric for $\displaystyle{SU(3) \over U(1)
\times U(1)}$ depends on two parameters. The coherent state techniques
yield the K\"ahler form in variables where the two parameters directly
correspond to the values $(n, m)$ used to specify the representation
above. In addition, the representations \gensol\ yield an embedding of
the two surfaces in $4$ and $6$ dimensions respectively. The specific
form of this embedding is also presented in Appendix A. As mentioned
in the Appendix, we see that in the limit when $m \rightarrow \infty$
and $n/m \rightarrow 0$, the $\displaystyle{SU(3) \over U(1) \times
U(1)}$ surface degenerates to $CP^2$.

\leftline{\it {4.3.2 \ \ Dipole Moments}}

We conclude this section with a discussion of the various dipole
moments and charges induced in the two cases. The $CP^2$ case is
presented in some detail first, the $\displaystyle{SU(3) \over U(1)
\times U(1)}$ results which can be obtained in the same way, are
discussed more briefly at the end. In the following discussion we set
$p=0$ for simplicity, so that we are dealing with a $D0$-brane.
Nothing essential changes in the $D1$-brane context. In the $CP^2$
case one expects the configuration \gensol\ to carry a dipole moment
for the $F^{(6)}$ and $F^{(4)}$ form field strengths and  a
$D0$-brane charge. The two dipole moments are determined by the
couplings:
\eqn\cpmom{\eqalign{& {T_p \over \lambda^2} \int dt\; F^{(6)}_{0ijklm}
Tr([X^i,X^j] [X^k,X^l] X^m]) \cr 
{\rm and} \quad & i {T_p \over \lambda} \int dt \;F^{(4)}_{0ijk} Tr([X^i,X^j]X^k) }} 
respectively \foot{Our conventions are that in the expression for the
$F^{(6)}$ brane dipole moment each distinct pair of commutators appears
only once. i.e.,
\eqn\newe{\eqalign{ F^{(6)}_{0ijklm}  Tr([X^i,X^j] [X^k,X^l] X^m]) =
F^{(6)}_{012345} \bigl ( & 
Tr([X^1,X^2] [X^3,X^4] X^5) - Tr([X^1,X^3] [X^2, X^4] X^5)  \cr
& +Tr([X^2,X^3] [X^4,X^5] X^1) + \cdots 
\bigr ) .}}}.
From the four-brane perspective, no charge for the $F^{(6)}$ field
strength arises because the $CP^2$ surface is embedded in $R^8$.  The
$D0$-brane charge and dipole moments arise due to wrapping a {\it
single} four-brane on $CP^2$ (to get $k$ four-branes one needs to
begin with a reducible representation containing $k$ copies of the
symmetric representation). A $U(1)$ field strength is turned on in the
world volume of the four-brane. It is
\eqn\fcpt{{\cal F}_{z_i \bar{z}_j} = K_{z_i \bar{z}_j},}
where $K_{z_i \bar{z}_j}$ the K\"ahler form for $CP^2$ is given in
(B.13) of Appendix B. This gauge field has a non-trivial first and
second Chern class \foot{ In fact $\pi_2(SU(3)/U(2)) = Z$, and
$\int_{S^2}{\cal F} = m$.}.

The $D0$-brane charge in the four-brane theory is given by
\eqn\chrgdz{N={T_{p+4} \over 2 T_{p}} \lambda^2 \int {\cal F} \wedge
{\cal F}.}
Evaluating the RHS gives an answer $m^2/2$ (for $m \gg 1$), this
agrees with the dimension of the representation, $N$, from \dimsut.

The dipole moments for $F^{(6)}$ and $F^{(4)}$  are given by the couplings:
\eqn\secdi{\eqalign{& T_{p+4} \int F^{(6)}_{0ijklm} X^m dX^i \wedge dX^j
\wedge dX^k \wedge dX^l \  dt \cr 
{\rm and} \quad     & T_{p+4}\  \lambda \int F^{(4)}_{0ijk}
dX^i \wedge dX^j  \wedge {\cal F} \  X^k \  dt.}} 
The integrals above are understood as being done over the $CP^2$
manifold.  In (B.15) we describe how $CP^2$ is embedded in $R^8$. This
embedding determines $X^m$ and the differential $dX^m$ as functions of
$(z_i, \bar{z}_i)$ and their differentials. Also, ${\cal F}$ is
determined as a function of $z_i$ from \fcpt.

One can show that \secdi\ agrees quantitatively (in the classical
limit, for large $m$) with \chrgdz\ for any perturbation $F^{(6)}$ and
$F^{(4)}$ \foot{Here we mean a perturbation about the background
\genf.}. To show this it is convenient to use the coherent state
basis, described in Appendix B to evaluate the trace in \cpmom, which
can be expressed as:
\eqn\cpmomt{\eqalign{& {T_p \over \lambda^2} {m^2 \over 4 \pi^2}
 \int {d^2z_1 d^2 z_2 \over (1+ |z_1|^2 + |z_2|^2)^3}   dt\; F^{(6)}_{0ijklm}
\langle z_i|[X^i,X^j] [X^k,X^l] X^m|z_i \rangle \cr
{\rm and} \quad & i {T_p \over \lambda} {m^2 \over 4 \pi^2} \int {d^2z_1 d^2 z_2 \over
 (1+ |z_1|^2 + |z_2|^2)^3} dt \;F^{(4)}_{0ijk} 
\langle z_i|[X^i,X^j]X^k|z_i \rangle }}
respectively. Now it is straightforward to show that the contribution
to \secdi\ and \cpmomt\ from each point in $CP^2$ as parametrized by
$(z_1, z_2)$ agree.  In fact, since $CP^2$ is a coset, all points on
it can be related by the action of the group $SU(3)$ and it is enough
to prove that these contributions agree at some one point in the
manifold.  The calculation is greatly simplified by choosing this
point to $z_1=z_2=0$, since the corresponding coherent state is then
the lowest weight state itself.  As an example consider the dipole
moment for $F^{(4)}$.  The contribution to \cpmomt\ from the vicinity
of this point (in the classical limit) is \eqn\conor{i{T_p \over
\lambda} {m^2 \over 4 \pi^2} d^2z\ d^2\omega F^{(4)}_{0ijk} \ \langle
lws|[X^i,X^j]|lws \rangle \ X^k(0).}  Using the $SU(3)$ algebra, one
can show that for a general $F^{(4)}$ this agrees with the
corresponding contribution in \secdi.

Before proceeding, let us make one parenthetical remark which will be
of relevance in the supergravity discussion of section 5. From (B.14),
(B.15), we see that at $z_1=z_2=0$ the coordinates $X^1, X^2, X^4,
X^5$ lie along the $CP^2$ surface. The contribution from the
neighborhood of this point to the dipole moments coupling to
$F^{(4)}_{012k}$ and $F^{(4)}_{045k}$ are in the ratio $\langle
lws|[X_1, X_2]|lws \rangle / \langle lws| [X_4, X_5]|lws \rangle$.
From the $SU(3)$ algebra and (B.14) we see that these are equal. In
the supergravity dual a $B$-field will be turned on in the vicinity of
the $(p+4)$-brane. This field has rank four and will be specified by
two parameters $b_1, b_2$. From the argument just given, one can argue
that $b_1/b_2=1$.

Similarly in the $\displaystyle{SU(3) \over U(1) \times U(1)}$ case,
we start with $D0$-branes for simplicity, and the configuration
\gensol\ gives rise to dipole moment for $F^{(8)}, F^{(6)}$ and
$F^{(4)}$. These can be calculated both from the perspective of the
$D0$-brane theory, in the form of traces over matrices as in \cpmom, \cpmomt,
and in the six-brane theory by couplings analogous to \secdi (once
again a $U(1)$ gauge field ${\cal F}$ given by the K\"ahler form (C.7), is
turned on in the world volume theory of the six-brane).  The resulting
answers agree quantitatively. In the supergravity dual the geometry
contains a $(p+6)$-dimensional brane. In the vicinity of this brane a
$B$-field is turned on which is characterized by three parameters
$b_1, b_2$ and $b_3$.  As in the $CP^2$ case, their ratios can be calculated
by considering the corresponding local contributions to the dipole
moments in the gauge theory. These are determined by the integers
$(n, m)$ which characterize the representation.

\subsec{Fuzzy $S^4$}
So far we have considered only $F^{(p+4)}$-form field strength
backgrounds. We end this section by considering one example of a
$F^{(p+6)}$-form field strength background. The full potential is now
\eqn\fullpotnew{V={T_p \over \lambda^2} \int d^{p+1}x \Bigl [-{1 \over 4} \sum_{a,b}
 Tr([X^a,X^b]^2) 
+ {1\over 5} F^{(p+6)}_{01...pijklm} Tr([X^i,X^j][X^k,X^l]X^m) \Bigr ] }
Setting
\eqn\valfsix{ F^{(p+6)}_{01...pijklm}=-{f \over \lambda} \epsilon_{ijklm} , \ \ 
\{i,j,k,l,m \} \in  \{p+1,p+2,p+3,p+4,p+5 \} } 
we have an extremum at
\eqn\ext{X^{i}= r \gamma^i, \ \  i \in  \{p+1,p+2,p+3,p+4,p+5 \} }
and,
\eqn\valrhere{ r=  {5\over 6} {\lambda \over f}. }
In \ext\ the $\gamma_i$'s denote the Gamma matrices of $SO(5)$ in an
$N$-dimensional representation. One can verify that the extremum
\ext\ is not a minimum, putting in an ansatz of the form \ext\ and
varying $r$ one finds a runaway direction as $r \rightarrow \infty$.
It was argued in \ref\Taylor{J. Castelino, S. Lee and W. Taylor,
Nucl.Phys. {\bf B526}, 334-350 (1998), {\tt hep-th/9712105}.} that by
taking the $\gamma^i$ to be in the symmetric product of the
four-dimensional representation one obtains a fuzzy generalization of
$S^4$. This construction is quite different from the coset
construction for fuzzy surfaces which is the main concern of this
paper.

\newsec{Supergravity Duals}

In this section we turn to constructing the supergravity descriptions
of the fuzzy coset vacua. The gravity background for the ${\cal N}=4$
theory is the near horizon geometry of an extremal $D$-brane.
Deforming the ${\cal N}=4$ theory corresponds to turning on additional
perturbations in this background. No-hair theorem considerations
suggest that such deformations give rise to singularities in
general. In fact, we will find that singularities, corresponding to
$(p+4)$- and $(p+6)$-dimensional brane sources, are present in the
gravity duals.  This makes the discussion below also of interest from
the point of view of studying singularities in gravity via their gauge
theory duals.

Our discussion follows \PS\ closely. These authors analyzed the
gravity dual for the fuzzy $S^2$ case and showed that the gravity
background contained $5$-branes in the interior. This establishes that
the infra-red behavior of the configuration \four\ \redrep\ is
governed by the $(5+1)$-dimensional $D5$-brane theory (at least at
large 't Hooft coupling). The strategy in \PS\ was to solve the
gravity equations by choosing parameters which allowed spacetime to be
divided into two region in which the stress-energy is dominated by the
$(p+2)$-form field strength and $(p+4)$-form fields strengths
respectively.  The first region is essentially a multi-centered
version of the $Dp$-brane geometry while the second is the
$D(p+2)$-brane metric. Denoting the two field strengths by $F^{(p+2)}$
and $F^{(p+4)}$ respectively, the first region corresponds to
$(F^{(p+4)})^2 /(F^{(p+2)})^2 \ll 1$ while the second to
$(F^{(p+4)})^2 /(F^{(p+2)})^2 \gg 1$.  The crossover region between
the two is described by the gravity background dual to the
$D(p+2)$-brane with a $U(1)$ (or equivalently NS B-field) turned on
along its world volume\foot{In the discussion below we will sometimes
refer to this background as the noncommutative brane geometry.},
\HI, \MR\  and \AJ. This has a region of validity that overlaps with both the $Dp$-
and $D(p+2)$-brane metrics.  Here we will follows the same strategy.
There will be one variation: the fuzzy surfaces correspond to
$Dp$-branes distributed on higher dimensional surfaces, accordingly
the spacetime will sometimes be divided into more than two regions and
branes of dimension $p+4$ and higher will enter the story as well. In
the discussion below the required conditions on parameters will be
found in a self-consistent manner. We assume a region of parameter
space exists giving rise to some solution, construct the solution in
parts, then deduce the required conditions on the parameters by
demanding consistent overlap between the different parts.

One limitation of our analysis, mentioned in the introduction, needs
to be pointed out here. In the solutions we construct, we show that
the noncommutative $(p+2)$-brane metric and the $p$-brane metric
overlap consistently only to leading order in the perturbation
\foot{Actually, as was mentioned above, higher dimensional field strengths enter
as well, but this is an inessential feature we suppress at the
moment.}, $(F^{(p+4)})^2/(F^{(p+2)})^2$. This is not enough to
establish that the solutions exist. It is particularly important in
the cases under discussion here to go further, because, as was
mentioned in section 3, all supersymmetries are broken. Two arguments
indicate that going to the next order in the perturbation should be
enough to establish the existence or lack thereof of these solutions.
In the gravity calculation, the radius of the surface is determined in
terms of the perturbation $F^{(p+4)}$ only at the quadratic order.
From the gauge theory perspective, since supersymmetry is broken, one
expects that the most probable cause for destabilizing these surfaces
is that the scalar fields acquire large masses - these are effects
quadratic in the perturbation \foot{In AdS/CFT correspondence, scalar
masses are dual to Kaluza-Klein harmonics different from $F^{(p+4)}$,
so if large mass terms are a concern, one might hope to stabilize the
fuzzy surfaces by turning on appropriate values for these other modes
as well. However, in the case of $AdS_5 \times S^5$, the traceless
mass terms have supergravity duals but the trace component is dual to
a string mode. Thus it is not clear that enough freedom available. For
the $D1$-brane geometry considered below, the map between sugra modes
and operators in the ${\cal N}=4$ theory is less well understood and we could
not settle if all masses can be adjusted in the supergravity
approximation.}. Unfortunately, the equations get rapidly complicated
beyond leading order and we have not been able to push the analysis
further. Accordingly the solutions we present here should be viewed as
only a first step in a more complete analysis \foot{We should note
that \PS\ does determine the radius of the two-sphere, which we
mentioned above was sensitive to quadratic effects, but not from the
gravity equations directly. Rather they consider a brane probe. Terms
quadratic in the perturbation play an important role but their
normalization can be determined by appealing to supersymmetry. In the
present context, the absence of supersymmetry comes in the way of
using this approach.}.

We turn now to constructing the gravity backgrounds. For appropriate
regions of parameter space, within the approximation mentioned above,
we will find gravity solutions corresponding to $S^2 \times S^2$ and the two
cosets of $SU(3)$. The $S^2 \times S^2$ example is considered in some
detail in the context of $AdS_5 \times S^5$.  The cosets of $SU(3)$,
which require eight transverse dimensions, are discussed more briefly
in the $D1$-brane background.

\subsec{Dual Description of fuzzy $S^2 \times S^2$}
The near horizon geometry of $D3$-branes is $AdS_5\times S^5$.  We are
interested in the theory obtained by turning on additional trilinear
terms \fform, \stst, in the ${\cal N}=4$ Lagrangian. In the AdS/CFT
correspondence a combination of this operator and the fermionic mass
term is dual to the three-form field strength \wittenads,
\ref\KRN{H. J. Kim, L. J. Romans, P. van Nieuwenhuizen,
Phys. Rev. {\bf D32}, 389 (1985).}, \ref\GM{M. Gunaydin, N. Marcus,
Class. Quant. Grav. {\bf 2}, L11 (1985).},
\eqn\defgt{G_3= F_3 - (C+i e^{-\phi}) H_3,}
where $F_3,H_3$ stand for the RR and NS three-form field strengths,
and $C, \phi$ for the axion and dilaton. Exactly, this case was
studied in \PS. The main difference here is that the required
perturbation \stst\ corresponds to turning on equal masses for the
four gauginos,
\eqn\massga{m_{\lambda_i}= f_1+ i f_2,}
and thus breaks supersymmetry completely \foot{In fact even in the
limit when say $f_2 \rightarrow 0$ and the surface reduces to a single
$S^2$ the background \stst\ does not preserve supersymmetry.}.

As mentioned above we choose parameters so that the spacetime can be
divided into distinct regions each dominated by one $p$-form field
strength. The effects of the additional perturbation die away close to
the boundary and the geometry in this region is always the
multi-centered $D3$-brane metric. We will see that deep in the
interior, corresponding to the far infra-red in the gauge theory, the
geometry is a $7$-brane one.  This leaves room for two possibilities,
both of which will be realized below.  When one of the two spheres has
a radius much smaller than the other, the $D3$-brane metric first goes
over into a multi-centered version of the noncommutative $5$-brane
metric. In turn, proceeding further along the radial direction, this
turns into the $7$-brane metric. On the other hand when the two radii
are more comparable, the $D3$-brane metric directly goes over to the
$7$-brane.  We analyze these two cases in turn below.

Before proceeding, let us relate the parameters in the gravity
solutions to those which appeared in section 4. \ststm, \ststma\
depend on four parameters, the strength of the perturbations $f_1,f_2$
and the size of the two $SU(2)$ representations $m,n$. From our
discussion in section 4.2, applied to the $p=3$ case, it follows that
\ststm, \ststma, correspond to taking one 7-brane, and in this case
$mn=N_3$ the number of 3-branes. Also, a $U(1)$ field is turned on in
the world volume of the 7-brane. The strength of this field is
determined by $m,n$.  On the gravity side we will find that the
solutions depend on the two radii $r_1,r_2$ and on two parameters
$b_1,b_2$ which specify the rank four NS B-field. The product,
$b_1b_2$ will be determined in terms of the number of 3- and 7-branes.

\leftline{\it { 5.1.1 \ \  $D3 \rightarrow D5 \rightarrow D7$}}

The multi-centered three-brane solution is
\eqn\mdt{\eqalign{ds^2 =&  H_3^{-1/2}\eta_{\mu\nu}dx^{\mu}dx^{\nu} +
H_3^{1/2}\sum_{m=1}^6 dy^{m}dy_{m}, \cr
                   e^\phi=&g_s, \cr
                      F_5=& d\chi_4 + *d\chi_4,  \cr
{\rm where} \quad  \chi_4= & { 1\over g H_3} dx^{0} \wedge dx^{1} \wedge dx^{2} \wedge dx^{3}.  }}

Here,  $H_3$, the harmonic function is
\eqn\hardt{\eqalign{H_3=&{ R_3^4 \over 16 \pi^2} \int d\Omega_1 d\Omega_2
{1 \over |\vec r - \vec r(\Omega_1,\Omega_2)|^4} \cr {\rm with }
\quad R_3^4=& 4 \pi g_s N_3 l_s^4 \cr {\rm and } \quad r^2 =& y^m y_m.
} }
$H_3$ corresponds to distributing the branes uniformly on the two $S^2$'s. We
take one sphere of radius $r_1$ to lie in the $y_1,y_2,y_3$ directions
and the second sphere of radius $r_2$ to lie in the $y_4,y_5,y_6$
directions.  Here, we also assume that, $r_2 \gg r_1$.

Besides the fields \mdt\ the three form $G_3$ is also turned on.
Asymptotically, as $r \rightarrow \infty$ this has the form:
\eqn\asyt{G_3=\alpha_3 r^{-4} + \beta_3 r^{-6}}
with $\alpha_3$ being the non-normalizable mode which is determined by
the coefficient of the operator that is turned on in the gauge theory,
and $\beta_3$ being the normalizable mode which corresponds to the
five brane dipole moment. In addition, although we do not explicitly
demonstrate it, the axion is also excited corresponding to the
quadrupole seven brane moment discussed in section 4.2.

Now, let us approach the point $(y_4,y_5,y_6)=(0,0,r_2)$ close to the
second sphere. Denote $(y_1,y_2,y_3)$ by $\vec y$ and define $\rho^2=
{\vec y}^2 + (y_6 -r_2)^2$ to be the distance in the directions
transverse to the two- sphere. For $r_1 \ll \rho \ll r_2 $ the
harmonic function takes the form:
\eqn\approxhar{H_3 \simeq {1 \over 16 \pi} {R_3^4 \over  r_2^2 } \int
d\Omega_1{1 \over [(w_3-r_2)^2 +|\vec{y}-\vec{y}(\Omega_1)|^2]} }
where the integral is over the two- sphere of radius $r_1$.

Next consider the geometry for $D5$-branes extending along $X_1, X_2,
X_3,X_4, X_5$ directions with an $NS$ B-field turned on in the $X_4,
X_5$ plane. We consider a multi-centered version of this geometry
where the $5$ branes are distributed in an $S^2$ of radius $r_1$
(lying in the $y_1,y_2,y_3$ directions)
\eqn\mnc{\eqalign{ds^2= H_5^{-1/2} [-dx_0^2 + dx_1^2 &+dx_2^2  + dx_3^2 
+ h(dx_4^2 + dx_5^2)] + H_5^{1/2}(dy^mdy_m), \cr
e^{\phi}= g_s H_5^{-1/2} h^{1/2}, \quad h^{-1}=& 1+{b^2 \over H_5 l_s^4},  
\quad  B_{45}= {b l_s^2 \over l_s^4 H_5 + b^2}, \cr
F^{(7)}_{012345r} = {1\over g_s} {l_s^2 \over b} h \  \partial_r H_5^{-1}, & \quad 
F^{(5)}_{0123r} =  {1\over g_s} \partial_r H_5^{-1}, }}
where the harmonic function $H_5$ is \foot{From our discussion of
dipole moments in section 4.2, it follows that in the gravity dual of  \ststm, \ststma,
$N_5 = m$, more generally, $N_5=N_7 m$.}
\eqn\harmany{H_5(\vec y)={ g_s N_5 b_2  \over 4 \pi} \int d\Omega_2
 {1 \over |\vec y-\vec y(\Omega_2)|^2}. }
The parameter $b_2$ is related to the strength of the NS B-field and
also determines $F^{(3)}$ and $F^{(5)}$ \foot{Once again, although we do not
describe it explicitly, the axion is also excited in the solution
\mnc.}.  It is easy to see that the effects of $F^{(5)}$ are negligible
when $b_2^2/(l_s^4 H_5) \ll 1$ and dominate over $F^{(3)},H_3$ when
$b_2^2/(l_s^4 H_5) \gg 1$. To compare with the multi-centered
$D3$-brane solution we therefore consider the region $b_2^2/(l_s^4
H_5) \gg 1$ in \mnc. In this region, $h \simeq H_5 l_s^4/ b_2^2$ and
the two metrics are the same. To see this identify $y_1,y_2,y_3$ in
the two metrics; $(y_6-r_2)$ in \mdt, with $y_4$ in \mnc. Finally,
identify $y_4,y_5$ in \mdt, with $x_4,x_5$ in \mnc, after a
rescaling.  The two metric then agree, provided,
\eqn\conda{{ N_5 \over \pi} {b_2 \over l_s^4} r_2^2 = N_3.} Similarly the dilaton
and $F^{(5)}$ also agree in this region. Verifying if the three-form
field strength matches involves a subtlety.  The three-form on the
3-brane side depends on an unknown parameter (which fixes the
normalizable mode in the asymptotic region). To determined both this
parameter and $b_2$ in terms of the non-normalizable mode's
coefficient requires us to work to next to leading order in the
perturbation $(G_3)^2/(F^{(5)})^2$.  As mentioned at the beginning of
this section, the second order calculation is beyond the scope of this
paper. Before proceeding let us comment on \conda. In the $5$-brane
theory the $3$-brane density is determined by $b_2/l_s^4$. Since the
$5$-brane is wrapped on a sphere of radius $r_2$ \conda\ follows.

We now go to smaller values of $|\vec y|$, away from the crossover
region in the 5-brane metric, \mnc. Once $b_2^2/(l_s^4 H_5) \ll 1$, $h
\simeq 1$, and the solution reduces to the multi-centered $5$ brane
with no world volume B-field.  Going to even smaller values of
$|\vec y|$ one finds that the stress energy in the axion field begins
to dominate, and \mnc\ in turn crosses over to the 7-brane
solution. The discussion of this transition is similar to the one
above so we will be brief. The only new feature is that the the
harmonic function in the 7-brane case varies logarithmically.  The
noncommutative 7-brane metric is given by
\eqn\msc{\eqalign{ ds^2 =& H_7^{-1/2}(dx_0^2 + \sum_{i=1, ..3} dx_i^2 +h_2(dx_4^2+dx_5^2)
                                                 + h_1(dx_6^2 + dx_7^2) ) \cr
                         & + H_7^{1/2}(d\rho^2 + \rho^2 d\phi^2) \cr
{\rm where} \quad H_7 =&  {C_7g_s b_1 b_2 \over l_s^4} N_7\ {\rm  ln }
\Bigl |{2 r_1 \over \rho} \Bigr | \cr {\rm and} \quad h_1^{-1} =& 1+
                         { b_1^2 \over H_7  l_s^4}. }}
In addition, the axion and the three-form fields, $H_3$ and $F^{(3)}$
are also excited.  $C_7$ above can be determined by ensuring that the
axion has the correct periodicity in $\phi$, for $N_7$ branes. After
appropriately changing variable one can show that \msc\ agrees with
the metric in
\mnc\ in the region $b_1^2 \gg H_7 l_s^4 $,
provided the condition
\eqn\condb{2 C_7  N_7 {b_1\over l_s^4}  r_2^2= N_5,}
is met.
\condb\ is analogous to \conda.

Let us summarize all the conditions required for solution described
above to be valid. For the metric \mdt\ and \mnc\ to be both valid in
the crossover region between the $5$-brane and $3$-brane solutions we
have:
\eqn\condc{r_1 \ll \sqrt{N_5 g_s \over b_2} l_s^2 \ll  r_2. }
Substituting for $b_2$ from \conda\ yields
\eqn\condd{r_1 \ll \sqrt{ g_s N_5^2 \over \pi N_3} r_2 \ll r_2.}
For \mnc\ and \msc\ to have a common region of validity in turn
implies (after dropping constants and taking the logarithm to be $\sim
1$):
\eqn\conde{{b_1 \over g_s N_7 b_2} \gg 1.}
Substituting for $b_1$, $b_2$ from \condb\ \conda\ gives
\eqn\condf{{1 \over g_s} {N_5^2 \over  N_7^2 N_3} r_2^2 \gg r_1^2 }
Finally, demanding that the curvature is small in string units except
at the seven brane singularity, gives:
\eqn\condg{r_2 < \sqrt{N_3 g_s} r_1.}
\condd, \condf, and \condg, are the independent constraints. A little
thought shows that for $g_s \ll 1, N_3 \gg 1$, $N_7 \sim O(1)$, they
can all be met by appropriately choosing $r_1,r_2$ and $N_5$ \foot{Or
equivalently, from \conda\ $r_1,r_2$ and $b_2$.}.

\leftline{\it { 5.1.2 \ \  $D3 \rightarrow D7$ : }}
We now turn to considering the second possibility which is realized
when the radii of the two $S^2$'s are more comparable. Here, the
$D3$-brane geometry directly goes over to the $7$ brane solution. In
this case let us consider the harmonic function \mdt\ in the vicinity
of the point $(y_1,y_2,y_3,y_4,y_5,y_6)) =(0,0,r_1,0,0,r_2)$.  Define
$\rho^2=(y_3-r_1)^2+(y_6-r_2)^2$, then for $\rho \ll r_1$ and $\rho
\ll r_2$ we have
\eqn\secformh{H_3={ \pi \over 4} {g_s N_3 l_s^4 \over r_1^2 r_2^2}\
{\rm ln} \Bigl [{4 r_1^2 r_2^2 \over ( r_1^2 +  r_2^2) \rho^2} \Bigr ].}

Compare this with the geometry for a 7-brane with a rank four
NS-field turned on in its world volume. This solution is given by
\foot{Once again the coefficient $C_7$ can be fixed by demanding
periodicity in $\phi$.}
\eqn\mstwo{\eqalign{ds^2 =& {\tilde H_7}^{-1/2}[-dx_0^2 +
\sum_{i=1,..3} dx_i^2 + h_1(dx_4^2+dx_5^2) + h_2(dx_6^2+dx_7^2) ] + \cr
& {\tilde H_7}^{1/2}(d\rho^2+\rho^2 d\phi^2),  \cr
{\rm where} \quad {\tilde H_7} =& {C_7g_s b_1 b_2 \over l_s^4} N_7
\ {\rm ln} \Bigl [{4 r_1^2 r_2^2 \over ( r_1^2 +  r_2^2) \rho^2} \Bigr ], \cr
{\rm and } \quad  h_i^{-1} =& 1+ {b_i^2 \over H_7 l_s^4}, \ i=1,2. }}

In the region where
\eqn\conddir{b_i^2 \gg H_7 l_s^4} the two metrics  \mstwo\ and \mdt, \secformh,
(as well as other fields like the dilaton, $F^{(5)}$ etc. which we do not
explicit exhibit) agree provided,
\eqn\condpb{{ 4 C_7 \over \pi} {N_7 r_1^2 r_2^2 b_1 b_2  \over l_s^8} = N_3. }
For \conddir to be true we have, approximating the logarithm by unity and
dropping various constants, that
\eqn\condpa{N_7 g_s \ll b_1/b_2 \quad  {\rm and} \quad N_7 g_s \ll b_2/b_1.}
Finally, the requirement that the curvature is small until one comes
close to the seven brane is met once $g_s N_3 \gg 1$.  In summary for
the solution \mstwo, \mdt, to exist, $r_1,r_2,b_1,b_2$ must thus
satisfy the conditions, \condpa\ and \condpb.

Let us end with two comments. First, one can use $S$-duality to
generate additional solutions, both for the case in section (5.12) and
here. These are valid in different (and somewhat complimentary)
regions of parameter space.  Second, as was mentioned at the outset of
this section, our analysis in the various crossover regions has been
to linear order in the perturbing RR potential.  We need to go beyond
this, at least to the next order, before conclusively establishing the
existence, or lack thereof, of these solutions. Since supersymmetry is
broken, it is quite likely, that such a second order analysis will
reveal that some of the solutions constructed here are unstable. But
hopefully, some will survive, yielding gravity backgrounds duals to
fuzzy surfaces.

\subsec{Gravity Duals to Cosets of $SU(3)$.}
We turn next to the cosets of $SU(3)$. In this case one needs at
least eight transverse dimensions. We will work with the $D1$-brane
system below. Holography is not as well understood in this context as
it is for the $D3$-brane system but this  is not a big limitation for our
limited analysis below. Our discussion will be somewhat brief, since
many of the essential points have been covered above.

We start with the $CP^2$ case then turn to $SU(3)/(U(1) \times U(1))$ in
the next section. In section 4.3 we saw that the irreducible
representation was completely determined by the dimension $N$, with
$m^2 = 2 N$. The discussion on the dipole moments (adapted for the
$D1$-brane system here) also showed that the irreducible representation 
corresponds to taking
the number of 5-branes, which wrap $CP^2$, $N_5=1$. To clarify the
origin of various terms, we keep $N_5$ as a free parameter below.
Also, here we will restrict ourselves to the simplest case where
spacetime gets divided into only two regions, one being the $D1$-brane
metric and the other the $D5$-brane geometry.

\leftline{\it{  5.2.1 \ \  $CP^2$}}
The multi-centered near-horizon limit of the $D1$-brane geometry is
given by:
\eqn\donemult{\eqalign{ds^2=& H_1^{-1/2}(-dx_0^2+dx_1^2) + H^{1/2}(\sum (dX^i)^2) \cr
e^{2 \phi}=& g_s^2 H_1, }} where $X_i$ denote the eight transverse
coordinates.  $H_1$ is the harmonic function which corresponds to
uniformly distributing the $D1$-branes over a transverse $CP^2$
surface.  From Appendix B  we have
\eqn\hone{\eqalign{H_1(\vec{ r})=& {1 \over 2 \pi^2} R_1^6 \int {d^2z_1
d^2z_2 \over (1+|z_1|^2+|z_2|^2)^3} {1 \over |\vec{r}-\vec{r}(z_1,
z_2)|^6} \cr R_1^6=& 32 \pi^2 g_s N_1 l_s^6.  }}
$\vec{r}$ is the eight-dimensional transverse vector, and
$\vec{r}(z_1,z_2)$ is determined by (B.14), (B.15).

We are interested in the deformed ${\cal N}=4$ theory, this is dual to
a background with a $F^{(5)}$ field strength perturbation turned
on. In addition as shown in section 4 a dipole moment for the
$F^{(7)}$ field strength is also expected, this is the magnetic dual
to $F^{(3)}$, so the $F^{(3)}$ field should also change from its value
in the ${\cal N}=4$ case. We do not exhibit these perturbations
explicitly below.

We now approach the point $z_1,z_2=0$ on $CP^2$.  At this point, we
see from (B.14), that only $X_3,X_8$ are non-zero.  Let
us denote their values to be $X^0_3, X^0_8$ repectively.  From (B.14)
we also see that at this point the coordinates parallel to the
surfaces are $X_1, X_2, X_3, X_4$. Define $\rho^2=(X_8-X^0_8)^2 +
(X_3-X_3^0)^2 + X_5^2+X_6^2$ to be the normal distance. Then, in the
vicinity of this point,
\eqn\limhone{H_1 = {16 \over 81} {R_1^6 \over R^4} {1 \over \rho^2},}
where $R$ is the radius of the surface, \rad.

In comparison the $D5$-brane metric with a rank four $B$-field turned
on is
\eqn\copardf{\eqalign{ds^2 = H_5^{-1/2}[-dx_0^2 + dx_1^2 +&
h_1(dx_2^2 + dx_3^2) + h_2(dx_4^2 + dx_5^2)] + H_5^{1/2}[\sum_{m=6}^9 dy_mdy^m], \cr
h_i^{-1} = 1+ {b_i^2 \over H_5 l_s^4}, \quad   H_5=& {R_5^2 \over r^2}, \quad 
R_5^2 = { g_s b_1 b_2 N_5 \over l_s^2}.   }}
Our discussion of $F^{(3)}$ dipole moments, section 4.3, implies that
\eqn\relbs{b_1=b_2 \equiv b.}
Now in the region where $b_i^2 \gg H_5 l_s^4$,
one can show, after a suitable
change of variables, that \donemult\ and \copardf\ agree provided,
\eqn\relsuta{{81 \over 512 \pi^2} {N_5 b^2 R^4 \over l_s^8} = N_1.}
\relsuta\ is analogous to \conda.

The other conditions which need to be met for this solution to be
valid are as follows. An overlapping region of validity for \copardf\
and \donemult, requires:
\eqn\fca{R^2 \gg g_s N_5 l_s^2.}
For the curvature to be small in string units except close to the
$5$-brane and for the string coupling to be small in the overlap
region requires:
\eqn\fcb{g_s N_1^{1/3} \ll {R^2 \over l_s^2} \ll g_s N_1. }
For $N_5 =1$, $g_s \ll 1$ and $N_1 \gg 1$ these can all be met.

\leftline{ \it { 5.2.2 \ \  $SU(3)/(U(1) \times U(1))$}}

Finally we discuss briefly the $SU(3)/(U(1) \times U(1))$ case. 
Here, the harmonic function \hone\ is replaced by the one
appropriate to distributing the branes on this coset. This can be
determined by  (C.5), in Appendix C. Approaching the point $z_i=0$ we
find that this metric crosses over to a seven-brane metric of the
form:
\eqn\metseven{\eqalign{ds^2= H_7^{-1/2}[-dx_0^2+dx_1^2+& h_1(dx_2^2 +
dx_3^2)+h_2(dx_4^2 + dx_5^2) + h_3(dx_6^2 + dx_7^2)] \cr & + H_7^{1/2}
(d\rho^2 +\rho^2 d \phi^2) \cr 
h_i^{-1}=1+ {b_i^2 \over H_7 l_s^4}, \quad   H_7=& {\tilde C_1}{g_s b_1 b_2 b_3 \over l_s^6} N_7 \  
{\rm ln } \Bigl [{{\tilde C_2} R^2 \over \rho^2} \Bigr ],  }} 
\eqn\conseven{{\rm provided} \quad {N_7 b_1 b_2 b_3 R^6 \over l_s^{12}} \sim N_1.}  
Here $R$ is the radius of the surface, defined in \genradius.
$\tilde{C}_{1,2}$ in \metseven\ are coefficients which can be
determined as in \mstwo. The metric \metseven\ depends on the
parameters $b_i$'s. The product is determined from \conseven\ in terms
of $R^2$. The ratios $b_i/b_j$ , we saw in our discussion in section
(4.3) are determined in terms of $m,n$ - the two integers specifying
the representation of $SU(3)$. There are additional conditions, for an
overlapping region of validity for \donemult\ \metseven, for the the
string coupling to be small and for the curvature to be small except
at the brane singularity.  When $m/n \sim O(1)$, these can all be met
for $ N_7 = 1$, $g_s \ll 1, N_1 \gg 1$.

\newsec{Acknowledgments}
We are grateful to A. Dhar and S. Wadia for illuminating conversations
regarding cosets, their coadjoint construction, and the use of
coherent states.  We also acknowledge discussions with A. Dabholkar,
S. Mukhi and S. Surya.

\appendix{A}{$SU(3)$ and its representations}

We follow the conventions of \ref\georgi{H. Georgi, {\it Lie Algebras
in Particle Physics: From Isospin to Unified Theories},
Benjamin/Cummings, Reading, Mass., 1982.}. The standard basis for the
defining (or fundamental) representation of $SU(3)$ consists of the
Gell-Mann matrices $\lambda_a, a=1, \cdots 8$, the generators being
$\lambda_a /2$. The {\it complex conjugate} (or anti-fundamental)
representation has generators $-\lambda^*_a /2$. We will denote the
gererators as $T^a$, and it will be clear from the context as to which
representation we are referring to.

The group $SU(3)$ has 2 simple roots, denoted by
\eqn\r{ \alpha^1 = (1/2, \sqrt{3}/2), \quad \alpha^2 = (1/2,
-\sqrt{3}/2).  
} 
where the first entry is the eigenvalue of $T_3$ and the second entry
the eigenvalue of $T_8$ in the adjoint representation. The 3 roots of
$SU(3)$ are $\alpha^1, \alpha^2$ and $\alpha^3 =\alpha^1 + \alpha^2$,
and the corresponding root vectors are denoted by $E_{\alpha^i}$ below.

The fundamental weights (with the above choice of $\alpha_i$'s) are 
\eqn\f{ \mu^1 = (1/2, 1/2 \sqrt{3}), \quad \mu^2 = (1/2, -1/2
\sqrt{3})
} 
The highest weight of the defining representation is $\mu^1$ while
that of the anti-fundamental representation is $\mu^2$.
The lowest weight vector of the fundamental representation is 
$-\mu^2$ and of the anti-fundamental representation is
$-\mu^1$.
 
The unitary irreducible representations (UIR's) of $SU(3)$ are
labelled by a pair of integers $(n, m)$. The states in the UIR $(n,
m)$ are constructed by taking the tensor product of $n$ fundamental
representations (i.e. $(1,0)$) and $m$ anti-fundamental
representations (i.e. $(0,1)$). The lowest weight state $|lws
\rangle_{(n,m)}$ for $(n, m)$ is 
\eqn\l {|lws \rangle_{(n,m)} = \underbrace{|lws \rangle_{(1,0)}\otimes
\ldots |lws \rangle_{(1,0)}}_{n\;\;{\rm factors}} \underbrace{|lws
\rangle_{(0,1)} \otimes \ldots |lws \rangle_{(0,1)}}_{m\;\;{\rm factors}} 
}  
All other states of $(n,m)$ can be obtained by by applying the
raising operators repeatedly.

\appendix{B}{$SU(3)$ coherent states}

We closely follow \Perelomov\ for notation and conventions in the
construction of coherent states. 

For any state $|\mu \rangle$ corresponding to a weight vector $\mu$,
the isotropy subgroup $H_{\mu}$ contains the Cartan subgroup $H = U(1)
\times U(1)$. For general weight vectors, $H_{\mu}$ coincides with
$H$, and the coherent state is characterized by the point of
${\cal M} =\displaystyle{{SU(3) \over U(1) \times U(1)}}$. However, $SU(3)$
also has {\it degenerate} representations for which the lowest weight
$\mu$ is singular, i.e., $\alpha. \mu =0$ for some root
$\alpha$. Consider for example $|lws \rangle_{(0,m)}$. In this case,
the isotropy group is not $U(1) \times U(1)$ but $U(2)$, and the
coherent state is characterized by a point of $\displaystyle{SU(3)
\over U(2)} = CP^2$.

To construct a coherent state, we start with the vector $|lws
\rangle_{(n,m)}$. This state satisfies $E_{-\alpha^i}|lws
\rangle_{(n,m)} = 0$, where $\alpha^i$'s are roots and $E_{-\alpha^i}$ are
the lowering operators. $SU(3)$ has 3 roots: $\alpha^1, \alpha^2$ and
$\alpha^3 = \alpha^1 + \alpha^2$. For both degenerate as well as
non-degenerate representations, the coherent state is defined as 
\eqn\c {|x \rangle = T(g) |lws \rangle,
} 
where $T(g)$ is the representation of the element $g \in
SU(3)$. Equivalently, It can also be written as
\eqn\e{ |\xi \rangle = N(\xi, \bar{\xi})
\exp{\Bigl [\sum_i \xi_i E_{\alpha^i} \Bigr ]} |lws \rangle_{(n,m)}
} 
where $E_{\alpha^i}$ are the raising operators. The normalization $N$ is
determined (up to an overall phase) by the condition 
\eqn\n{
\langle \xi|\xi \rangle =1.
}

\subsec{Coherent states corresponding $CP^2$}

We start with the $(0,m)$  representation of $SU(3)$. The
raising/lowering operators can be defined as 
\eqn\r { {T_1 \pm iT_2 \over \sqrt{2}} \equiv E_{(\pm
1,0)} \equiv T_{\pm}
}
\eqn\r { {T_4 \pm i T_5 \over \sqrt{2}} \equiv E_{(\pm
1/2,\pm \sqrt{3}/2)} \equiv U_{\pm}
}
\eqn\r { {T_6 \pm iT_7 \over \sqrt{2}} \equiv E_{(\mp
1/2,\pm \sqrt{3}/2)} \equiv V_{\pm}
}

Using the lowest weight state $|lws \rangle$ we can construct coherent
states
\eqn\c {|\xi_1, \xi_2 \rangle = N e^{\xi_1 T_{+} + \xi_2 U_{+}}
|lws \rangle \equiv | \xi_i \rangle
} 
where $T_{\pm}$ are the usual raising/lowering operators. The
normalization $N$ can be determined (up to an overall phase) to be
\eqn\n{|N|=\Bigl [ 1 + {|\xi_1|^2 + |\xi_2|^2 \over 2} \Bigr ]^{-m/2}
} 
Using the one-form ${\cal A}$ defined as
\eqn\a{\eqalign{{\cal A} =& \langle \chi_1, \chi_2|d|\xi_1, \xi_2
\rangle|_{\chi_i = \xi_i} \quad {\rm where} \cr
d|\xi_1, \xi_2 \rangle =& \partial_{\xi_i}|\xi_1, \xi_2 \rangle d\xi_i
+ \partial_{\bar{\xi}_i}|\xi_1, \xi_2 \rangle d\bar{\xi}_i. 
}}
we can calculate the K\"ahler form 
\eqn\K{K=d{\cal A}.}
To simplify equations, we define $z_i = \xi_i/ \sqrt{2}$ and work with
$z_i$'s henceforth. Then we find that 
\eqn\a{{\cal A} = d\;{\rm ln} N + \langle \xi_i|T_{+}|\xi_i \rangle d\xi_1 + \langle
\xi_i|U_{+}|\xi_i \rangle d\xi_2} 
\eqn\a{=d\;{\rm ln} N + {m \bar{z}_1 \over 1 + z_i \bar{z}_i} dz_1 + {m
\bar{z}_2 \over 1 + z_i \bar{z}_i} dz_2.}

Then the K\"ahler form $K$ is
\eqn\Kc{K=d{\cal A}={m \over (1 + z_k \bar{z}_k)^2} \Bigl [(1 + z_k \bar{z}_k)
\delta_{ij} - z_i \bar{z}_j \Bigr ] d\bar{z}_i \wedge dz_j}

We will also need the following:
\eqn\otherc{\eqalign{
\langle z_i|T_+|z_i \rangle = (m/2){\sqrt{2} \bar{z}_1 \over 1 + z_i
\bar{z}_i}\;,& \quad  
\langle z_i|T_-|z_i \rangle = (m/2){\sqrt{2} z_1 \over 1 + z_i \bar{z}_i}\;, \cr
\langle z_i|U_+|z_i \rangle= (m/2){\sqrt{2} \bar{z}_2 \over 1 + z_i
\bar{z}_i}\;,& \quad 
\langle z_i|U_-|z_i \rangle = (m/2){\sqrt{2} z_2 \over 1 + z_i
\bar{z}_i}\;, \cr  
\langle z_i|V_+|z_i \rangle = (m/2){\sqrt{2} \bar{z}_1 z_2 \over 1 +
z_i \bar{z}_i}\;,& \quad  
\langle z_i|V_-|z_i \rangle = (m/2){\sqrt{2} z_1\bar{z}_2 \over 1 +
z_i \bar{z}_i}\;, \cr 
\langle z_i|T_3|z_i \rangle = (m/2){(|z_1|^2 -1) \over 1 + z_i
\bar{z}_i}\;,& \quad
\langle z_i|T_8|z_i \rangle = {(m/2)\over \sqrt{3}} {(2|z_2|^2 -
|z_1|^2 -1) \over 1 + z_i \bar{z}_i}\;. 
}}

Let us end with one comment. \gensol, shows that the coordinates
$X^{p+i}$ are proportional to the generators $T^i$.  From \otherc\ we
then learn that $CP^2$ is embedded in $R^8$ as follows:
\eqn\embed{X^{p+a}(z_i)=f \langle z_i|T^a|z_i \rangle, a =1, ..8.}

\appendix{C}{Coherent States for $\displaystyle{{SU(3) \over U(1) \times U(1)}}$} 

Again, start with the lowest weight state $|lws \rangle \equiv |0
\rangle$ of an arbitrary UIR $(n, m)$ of $SU(3)$. The
coherent state is defined as
\eqn\coh{
|\xi\rangle = N e^{\sum \xi_i E_{\alpha^i}} |lws \rangle.
} 
It is easy to show that 
\eqn\re{
\exp{\xi_i E_{\alpha^i}} = \exp{\xi_1 E_{\alpha^1}}. \exp{(\xi_2 +
{\xi_1 \xi_3 \over 2 \sqrt{2}}) E_{\alpha^2}}. \exp{\xi_3 E_{\alpha^3}}
}
To minimize irritating factors of $\sqrt{2}$, we define $z_1 = {\xi_1
\over \sqrt{2}}, z_2 = {\xi_2 \over \sqrt{2}} + {\xi_1 \xi_3 \over 4}$
and $z_3 ={\xi_3 \over \sqrt{2}}$. 

We grind through to find the following quantities:
\eqn\n{
|N|^{-2} = \Bigl [{{1 + |z_1|^2 + |z_2 - z_1 z_3|^2} \over {1 + |z_2|^2 +
|z_3|^3}}\Bigr ]^{n+m}. [1 + |z_2|^2 + |z_3|^2]^m.
}
For notational simplicity, we define
\eqn\other{
A(z_i, \bar{z_i}) = 1 + |z_1|^2 + |z_2 - z_1 z_3|^2, \quad
B(z_i, \bar{z_i}) = 1 + |z_2|^2 + |z_3|^2.
}
We also need the following expectation values:
\eqn\other{\eqalign{
\langle z_i|T_1|z_i \rangle =& {(m+n) \over 2} \Bigl [{{\bar{z}_1
+ z_1} \over A(z_i, \bar{z}_i)} \Bigr ] - {n \over 2} \Bigl [{z_3
\bar{z}_2 + z_2 \bar{z}_3 \over B(z_i, \bar{z}_i)} \Bigr ],  \cr
\langle z_i|T_2|z_i \rangle =& {(m+n) \over 2i} \Bigl
[{{\bar{z}_1 - z_1} \over A(z_i, \bar{z}_i)} \Bigr ] - {n \over 2i}
\Bigl [{z_3 \bar{z}_2 - z_2 \bar{z}_3 \over B(z_i, \bar{z}_i)} \Bigr ], \cr
\langle z_i|T_3|z_i \rangle =& {(m+n) \over 2} \Bigl
[{|z_1|^2 -1 \over A(z_i, \bar{z}_i)} \Bigr ] - {n \over 2}
\Bigl [{|z_2|^2 - |z_3|^3 \over B(z_i, \bar{z}_i)} \Bigr ], \cr
\langle z_i|T_4|z_i\rangle =& {(m+n) \over 2} \Bigl [{{(z_2 - z_1z_3) +
(\bar{z}_2 - \bar{z}_1 \bar{z}_3)} \over A(z_i, \bar{z}_i)} \Bigr ] -
{n \over 2} \Bigl [{z_2 + \bar{z}_2 \over B(z_i, \bar{z}_i)} \Bigr ],  \cr
\langle z_i|T_5|z_i\rangle =& {(m+n) \over 2i} \Bigl
[{{(\bar{z}_2 - \bar{z}_1 \bar{z}_3) - (z_2 - z_1z_3)} \over A(z_i,
\bar{z}_i)} \Bigr ] - {n \over 2i} \Bigl [{\bar{z}_2 - z_2 \over
B(z_i, \bar{z}_i)} \Bigr ],  \cr
\langle z_i|T_6|z_i \rangle =& {(m+n) \over 2} \Bigl [{z_1
\bar{z}_2 + z_2 \bar{z}_1 + |z_1|^2(\bar{z}_3 + z_3) \over A(z_i,
\bar{z}_i)} \Bigr ] + {n \over 2} \Bigl [{\bar{z}_3 + z_3 \over B(z_i,
\bar{z}_i)} \Bigr ],  \cr
\langle z_i|T_7|z_i\rangle =& {(m+n) \over 2i} \Bigl [{z_2
\bar{z}_1 - z_1 \bar{z}_2 + |z_1|^2(\bar{z}_3 - z_3) \over A(z_i,
\bar{z}_i)} \Bigr ] +  {n \over 2i} \Bigl [{\bar{z}_3 - z_3 \over
B(z_i, \bar{z}_i)} \Bigr ],  \cr
\langle z_i|T_8|z_i\rangle =& {(m+n) \over 2\sqrt{3}} \Bigl
[ {2|z_2 - z_2 z_3|^2 - (|z_1|^2 -1) \over A(z_i,
\bar{z}_i)} \Bigr ] +  {n \over 2 \sqrt{3}} \Bigl [{|z_2|^2 + |z_3|^3
- 2 \over B(z_i, \bar{z}_i)} \Bigr ].  \cr
}}

To calculate the K\"ahler form, we also need 
\eqn\l{|N|^2 \langle 0|e^{\bar{\xi}_i E_{-\alpha^i}} e^{\xi_i
E_{\alpha^i}} E_{\alpha^3}|0 \rangle = -n {{\bar{z}_3 + z_1 \bar{z}_2} \over
B(z_i, \bar{z}_i)}
} 

Putting it all together to calculate the K\"ahler form, we get
\eqn\K{\eqalign{K =& {m+n \over A}(d\bar{z}_1\wedge dz_1 + d\bar{z}_2
\wedge dz_2) - {m+n
\over A^2} dA \wedge (\bar{z}_1 dz_1 + (\bar{z}_2 - \bar{z}_1\bar{z}_3)dz_2) \cr
+& {n \over B^2} dB \wedge (z_3 \bar{z}_2 dz_1 + \bar{z}_2 dz_2) - {n \over B}
(z_3 d\bar{z}_2 \wedge dz_1 + \bar{z}_2 dz_3 \wedge dz_1 + d\bar{z}_2
\wedge dz_2) \cr
-& {n \over B} (d\bar{z}_3 \wedge dz_3 + z_1 d\bar{z}_2 \wedge dz_3 +
\bar{z}_2 dz_1 \wedge dz_3)
+ {n \over B^2}(\bar{z}_3 + z_1 \bar{z}_2) dB \wedge dz_3.
}}

It is easy to see that in the limit $m \rightarrow \infty, n/m
\rightarrow 0$, (and doing a coordinate redefinition $(z_2 - z_1 z_3)
\rightarrow z_2$) \other\ go over into \otherc\ , and \K\ goes over to
into \Kc. Thus in this limit, $\displaystyle{SU(3) \over  U(1) \times
U(1)}$ degenerates to $CP^2$.

Finally, as in the $CP^2$ case above we can express $X^{p+a}$ \gensol,
as a function of the coordinate $z_i$ as
\eqn\embedsec{X^{p+a}=\langle z_i|T^a|z_i \rangle.}

\listrefs
\end